\documentclass[11pt,a4paper]{article}
\usepackage{ifpdf}
\usepackage{ae}
\usepackage[T1]{fontenc}
\usepackage[ansinew]{inputenc}
\usepackage{mathrsfs}
\usepackage{amsmath}
\usepackage{amssymb}
\pdfoutput=1
\usepackage{amsmath,amssymb,amsfonts,a4wide,graphicx,bm,times,psfrag,wrapfig,sidecap}
\usepackage{cite}
\usepackage[colorlinks=true,linkcolor=black, citecolor=black,urlcolor=black]{hyperref} 
\numberwithin{equation}{section}
\makeatletter \let\old@startsection=\@startsection
\renewcommand{\@startsection}[6]
{\old@startsection{#1}{#2}{#3}{#4}{#5}{#6\mathversion{bold}}}
\def\be{\begin{eqnarray}  }
\def\ee{\end{eqnarray}}
\newcommand\lab[1]{\label{eq:#1}}
\newcommand\nonu{\nonumber}
\def\vp{\varphi}

\def\hf{ {\textstyle{1\over 2}} } 

\begin{document}

\thispagestyle{empty}

\begin{flushright}

\end{flushright}

\vspace{1cm}
\setcounter{footnote}{0}

\begin{center}

{\Large\bf CLASSICAL AND QUANTUM TWO-BODY PROBLEM IN GENERAL 
RELATIVITY\footnote{Revised version of Trieste preprint IC/80/124 
(August 1980)}.}

\bigskip
  
 {\it Letters in Mathematical Physics 5 (1981) 359-366. 
 0377-9017/81/0055-0359 \$00.80.                                     
 
Copyright  1981 by D. Reidel Publishing Company.}
 
% \vspace{20mm} 
\vspace{16mm}

 A. MAHESHWARI, \footnote{Permanent address: Department of Physics, 
Regional College of Education, Mysore-6, 570006, India. }
  
{\it International Centre for Theoretical Physics, Trieste, Italy }

\bigskip
 
E.R. NISSIMOV$^{3}$  and I.T. TODOROV  
\footnote{Permanent address: Institute of Nuclear Research and 
Nuclear Energy, Bulgarian Academy of Sciences, Boul. Lenin 72, 1184 Sofia, 
Bulgaria.}\\

{\it International Centre for Theoretical Physics, Trieste, Italy
and Scuola Internazionale Superiore di Studi Avanzati, Trieste, Italy}
\end{center}

\vskip16mm

\noindent{ABSTRACT. 
The two-body problem in general relativity is reduced to the problem of an 
effective particle (with an energy-dependent relativistic reduced mass) in 
an external field. The effective potential is evaluated from the Born 
diagram of the linearized quantum theory of gravity. It reduces to a 
Schwarzschild-like potential with two different `Schwarzschild radii'. 
The results derived in a weak field approximation are expected to be 
relevant for relativistic velocities.
}

%   
%\newpage
%
\setcounter{footnote}{0}

\bigskip

\bigskip

% \bigskip

{\bf 1.} In both non-relativistic and special relativistic mechanics, classical 
and quantum, the two-body problem for (spinless) point particles is reduced 
to the conceptually simpler problem of a single effective particle moving 
in an external field. The only exception to this picture so far seems to
be the general theory of relativity, where the two-body problem has been 
treated in a considerably more complicated way: as a field-theoretic problem
with singularities [1, 2] (or as a problem of finite size bodies interacting
with a gravitational field [31).
Here we propose to treat gravitational two-particle interaction in much 
the same way as electromagnetic interactions have been tackled previously 
[4, 5] in the quasipotential approach [6] which found its natural place in 
the constraint Hamiltonian framework of References [7] and 
[8] .\footnote{A similar approach is being developed by a number of authors 
(see, e.g., References [9] and [10]); the 
reader will find a comprehensive bibliography in References [11] and [8].}
Unlike other first-order (in $1/c^2$) semi-relativistic treatments (based on 
a quantum field theoretic derivation of the two-particle potential) [12] ,
our approach is fully relativistic. Here we shall consider the two-body 
problem in the leading order of perturbation theory in $G$, the 
Newtonian gravitational constant. It is reduced to the problem of an 
effective particle (with an energy-dependent relativistic reduced mass) 
in an external Schwarzschild-like field with two 
different `Schwarzschild radii', in $g_{00}$ and $g_{ij}$  respectively.
 
\bigskip

\def\eff{\mathrm{eff}}
\def\pp{\mathbf{p}}
\noindent 
{\bf 2.} We shall briefly summarize the constraint Hamiltonian approach to 
the relativistic two-body problem and will introduce the notion of an 
effective particle in this approach.

We define the generalized two-point (spinless) particle mass shell as a 
14-dimensional sub-manifold of the 16-dimensional 'large phase space' 
$\Gamma$  of Minkowski space co-ordinates $x_l, x_2$ and  
four-momenta $p_1, p_2$, given by two first-class constraints. 
We postulate (as in [4, 5]) that the  difference $p_1^2 - p_2^2$   
is independent of the interaction:
\be
% \label{eqone}
\vp = \hf       (m_1^2 + p_1^2 - m_2^2 - p_2^2) = pP=0,
\lab{1}
\ee
where  $m_1, m_2$ are the masses of the two particles, $P$ and $p$ are 
the total and the relative momenta: 
\begin{align}
&P=p_1+p_2 , \quad
p=\mu_1 p_2 - \mu_2 p_1, \quad \mu_1+\mu_2=1, 
\nonu\\
&\mu_1-\mu_2= {m_1^2 - m_2^2\over w^2},
\quad
w^2 = - P^2 \ (>0).
%\label{eqdve}
\lab{2}
\end{align}
(We are using the space-like signature $-+++$  for the metric tensor.)

The non-relativistic reduced mass m is defined by the equation 
$mM = m_1m_2$, where $M = m_1 + m_2$ is the total mass. We use the same 
equation to define the relativistic reduced mass $m_w$, just 
replacing $M$ by the total relativistic mass $ w (= (-P^2)^{1/2})$:
\be
\label{eqtri}
m_w= {m_1 m_2\over w}.
\lab{3}
\ee
The effective particle four-momentum $P_{\mathrm{eff}}$ is then defined in 
the centre-of-mass frame (in which $P = (w, \mathbf{0}), p = (0,\mathbf{p})$) 
by
\be
\label{eqchetiti}
p_\eff = (E, \pp), \quad E= (m_w^2 + b^2 (w))^{1/2} = 
{w^2 - m_1^2 - m_2^2\over 2w},
\lab{4}
\ee
where $b^2(w)$ is the one-shell value of the relative momentum square
\be
b^2(w)={w^4 - 2 (m_1^2+m_2^2)w^2 +(m_1^2 - m_2^2)^2\over 4 w^2}.
\lab{5}
\ee
In the first approximation in the coupling constants (charges) $e_{1,2}$ , 
the electromagnetic interaction of two charged particles has been given by 
the Hamiltonian constraint [4, 5, 8]
\begin{align}
& H_{\mathrm{Coul} }= \hf[m_w^2 +\pp^2 - (E- V_{\mathrm{Coul}})^2]=0,
\quad  \pp^2 = p^2 = p_\eff^2  +{(Pp_\eff)^2\over w^2},
\nonu
\\
& V_{\mathrm{Coul}} = {e_1e_2\over 4\pi r}, \quad 
r= (x_\perp^2)^{1/2} = \left( x^2 + {(xP)^2 \over w^2}\right)^{1/2},
\quad x= x_1-x_2.
\lab{6}
\end{align}
(Note that the constraint (6) is manifestly a Poincare invariant; 
no semi-relativistic approximation of the type of the $1/c^2$ expansion 
has been made.) The idea of the present note is to describe in a 
similar fashion the gravitational interaction of two relativistic 
masses by setting
\be
H= H_{\mathrm{Grav}} = \hf[ m_w^2 + g^{\mu\nu} P_{\eff\,\mu} P_{\eff\,\nu}]=0,
\lab{7}
\ee
where $g^{\mu\mu}$  is some appropriate modification of the 
Schwarzschild metric.

\bigskip

\noindent
{\bf 3.} The actual computation of the electromagnetic Hamiltonian constraint 
(which includes corrections to $H_{\mathrm{Coul}}$) has been effected in 
the quasipotential approach to quantum electrodynamics [4, 5] . 
We shall pursue here a similar path starting with a standard linearized 
form of quantum gravity (cf. References [13]).

According to Dirac's general theory [14] , the quantum counterpart of the
first-class constraint (7) is the relativistic `Schr{\"o}dinger equation'
\be
\bigl\lbrack m_w^2 + \frac{1}{6}R - |g|^{-1/2} \partial_\mu
(|g|^{1/2} g^{\mu\nu}) \partial_\nu\bigr\rbrack \Psi = 0
\lab{8}
\ee
for the state vector $\Psi (x)$. Here $R$ is the scalar curvature. 
(As pointed out by Penrose [15] , the $R/6$ term is necessary in order 
to ensure conformal invariance of the zero mass limit.) [The 
Laplace-Beltrami operator provides the appropriate generally covariant 
ordering of the canonical variables 
$x_{\rm eff} (= -(Px) w^{-2}P^\mu + x^\mu_\perp)$ and 
$p_{{\rm eff}\,\mu} = - i\partial_\mu$.]    
The momentum space counterpart of (8) is to be identified with the local 
quasipotential equation [4, 5] (written here in the centre of the 
mass frame)
\begin{align}
& G^{-1}_w (\mathbf{p})\, {\widetilde \Psi}(\mathbf{p}) + 
(V{}_{*}{\widetilde \Psi})(\mathbf{p})
\nonu
\\
& \equiv 2w \lbrack \mathbf{p}^2 - b^2 (w)\rbrack {\widetilde \Psi}(\mathbf{p})
+ \int V(\mathbf{p},\mathbf{q}){\widetilde \Psi}(\mathbf{q})\frac{d^3 q}{(2\pi)^3} =0 \; , \;
\lab{9}
\end{align}
the potential $V$ is determined order by order in $G$ from the 
Lippmann--Schwinger-type equation:
\be
T + V + V {}_{*} G_wT = 0 \;\; ,\quad
G_w(\mathbf{k}) = \lbrack 2w(\mathbf{k}^2 - b^2(w)-i0 \rbrack^{-1}
\lab{10}
\ee
and from the Feynman expansion of the scattering amplitude 
$T = T_w(\mathbf{p},\mathbf{q})$ in a quantum theory of gravitationally 
interacting scalar particles.

We shall treat Equations (8) and (9) in the leading order approximation 
of perturbation theory. The linearized form of (8) is obtained by setting
\be
g_{\mu\nu} = \eta_{\nu\mu} + h_{\nu\mu} \;\;, \quad
\eta_{\nu\mu}=
\mathrm{diag}(-1,+1,+1,+1)\;\;,\quad |h_{\nu\mu}|\ll 1 \; ,
\lab{11}
\ee
and using $g^{\mu\nu} \approx \eta^{\nu\mu} - h^{\nu\mu}\;,\; 
|g| \approx 1 + h^\mu_\mu\; ,\; 
R \approx \partial_\mu \partial_\nu h^{\mu\nu} - \Box h^\mu_\mu$ 
(where Lorentz indices are raised and lowered by $\eta$). 
Up to terms of order $0(h^2)$ Equation (8) reads:
\be
\Bigl\{ m_w^2 - \Box + \bigl\lbrack h^{\mu\nu}\partial_\mu \partial_\nu
+ \frac{1}{6}(\partial_\mu \partial_\nu h^{\mu\nu}) +
(\partial_\mu h^{\mu\nu} - \frac{1}{2} \partial^\nu h^\lambda_\lambda)
\partial_\nu - \frac{1}{6}(\Box h^\mu_\mu)\bigr\rbrack \Bigr\}\Psi = 0
\lab{12}
\ee

Thus, in the leading order of perturbation theory we have:
\begin{align}
& {\widetilde h}^{\mu\nu}(p-q) q_\mu q_\nu +
\frac{1}{6}(p-q)_\mu (p-q)_\nu {\widetilde h}^{\mu\nu}(p-q) 
\nonu \\
& + \bigl\lbrack (p-q)_\mu {\widetilde h}^{\mu\nu}(p-q) -
\frac{1}{2} (p-q)^\nu {\widetilde h}^{\lambda}_{\lambda}(p-q)\bigr\rbrack
q_\nu  
\nonu \\
& - \frac{1}{6}(p-q)^2 {\widetilde h}^{\mu}_{\mu}(p-q) = 
\frac{1}{2w} T^{(1)}_w (\mathbf{p},\mathbf{q}) \; , 
\lab{13} \\
& p = (p_\mu) = (-E,\mathbf{p}) \;\; ,\;\; q = (-E,\mathbf{q}) \; ,
\nonu
\end{align}
where $E$ is given by (4) (and $\mathbf{p}^2 = \mathbf{q}^2 = b^2 (w)$
(on the mass shell)).

\bigskip

{\bf 4.} The Born approximation $T^{(1)}_w$ for the two-particle scattering 
amplitude is derived from the Lagrangian density:
\be
{\mathcal L}= - |g|^{1/2}\Bigl\lbrack \frac{R}{16\pi G}
+ \frac{1}{2} \sum_{k=1,2}
\bigl( g^{\mu\nu} \partial_\mu\Phi_k \partial_\nu\Phi_k + m_k^2 \Phi_k^2
+ \frac{1}{6}R \Phi_k^2\bigr)\Bigr\rbrack
\lab{14}
\ee
in the weak field approximation (11).\footnote{The naive $G$-perturbation 
theory of (14) is nonrenormalizable. According to the general discussion in
Reference [16], Equation (14) gives a correct description of gravitational
interactions only on tree-graph level and at a relatively low energy scale 
(much less that $10^19$ GeV for elementary particles). 
In order to compute $V$ consistently to arbitrary orders in $G$ from 
Equation (10) one should use a nontrivial renormalizable extension of 
(14) if there is any (at present only extended supergravity is a 
hopeful candidate).}
The expression (14) differs by the $\frac{1}{6}R \Phi_k^2$ term from the 
Lagrangian used in References [13] (corresponding to the $R/6$ in the 
Schr{\"o}dinger Equation (8)).
The one-graviton exchange diagram between particles 1 and 2 gives
\begin{align}
& T^{(1)}_w (\mathbf{p},\mathbf{q}) = 4\pi G \Gamma^{(1)}_{\kappa\lambda}
{\widetilde {\mathcal D}}^{\kappa\lambda,\mu\nu}(p^{(1)} - q^{(1)})
\Gamma^{(2)}_{\mu\nu} \; ,
\lab{15} \\
& \Gamma^{(k)}_{\mu\nu} = i[ p^{(k)}_\mu q^{(k)}_\nu +
p^{(k)}_\nu q^{(k)}_\mu - \eta_{\mu\nu}\bigl( p^{(k)}q^{(k)} 
+ \frac{1}{3}( p^{(k)}- q^{(k)})^2 + m_k^2 \bigr)
\nonu \\
& + \frac{1}{3}( p^{(k)}- q^{(k)})_\mu ( p^{(k)}- q^{(k)})_\nu ] 
\;\; , \;\; k=1.2\; ,
\lab{16} \\
& {\widetilde {\mathcal D}}^{\kappa\lambda,\mu\nu}(k) =
\frac{\eta^{\kappa\mu}\eta^{\lambda\nu} + \eta^{\kappa\nu}\eta^{\lambda\mu}
- \eta^{\kappa\lambda}\eta^{\mu\nu}}{k^2 - i0} \; ,
\lab{17}
\end{align}
\be
p^{(1)}=(E_1,\mathbf{p}) \;, \;\; p^{(2)}=(E_2,-\mathbf{p}) \;, \;\;
q^{(1)}=(E_1,\mathbf{q}) \;, \;\; q^{(2)}=(E_2,-\mathbf{q}) \;, \;\;
E_k = \mu_k w \; .
\lab{18}
\ee
Inserting (16), (17) and (18) into Equation (15), we obtain:
\be
T_w^{(1)} (\mathbf{p},\mathbf{q}) = 16\pi G \left\lbrack
\frac{2E^2 w^2 - m_1^2 m_2^2}{(\mathbf{p}-\mathbf{q})^2} - E w
- \frac{m_1^2 + m_2^2}{6} + \frac{1}{12}(\mathbf{p}-\mathbf{q})^2 
\right\rbrack \; .
\lab{19}
\ee

\bigskip
 
{\bf 5.} The next step is to evaluate $h_{\mu\nu}$ from Equations (13) and 
(19). To this end we shall use the Euclidean invariant 'stationary gauge' 
in which
\be
h_{0i}=0\;,\;\; h_00 = \frac{r_t}{r}\;\; (r_t = {\rm const})\;, \;\;
h_{ij} = B(r) x_i x_j \;\; ({\rm for} ~x \neq 0)\; .
\lab{20}
\ee
(The last condition means that we require the angular part of $ds^2$ 
to have its flat space form $r^2 (d\theta^2 + \sin^2 \theta d\varphi^2)$,
which is the standard co-ordinate choice for the Schwarzschild solution.) 
This amounts to setting
\begin{align}
&
h_{00}(\mathbf{p}-\mathbf{q}) = \frac{4\pi r_t}{(\mathbf{p}-\mathbf{q})^2}
\nonu \\
&
h_{ij}(\mathbf{p}-\mathbf{q}) = 
4\pi r_s \frac{(\mathbf{p}-\mathbf{q})^2\delta_{ij} -
2 (p_i-q_i)(p_j-q_j)}{(\mathbf{p}-\mathbf{q})^4} + C \delta_{ij} \; ,
\lab{21}
\end{align}
where $r_t, r_s$ and $C$ are constants of the motion. Inserting (19) and (21)
into Equation (13), we find \footnote{*The expression for $r_s$ does not 
coincide with the correct semi-relativistic approximation of Reference [1]. 
The results of Reference [1,2] indicate that the agreement will be restored
if one takes into account the semi-relativistic contribution to the 
effective potential coming from the Feynman diagrams of order $G^2$.}
\begin{align}
&
r_t = 2Gw\left\lbrack 1 - 
\frac{4b^2}{m_w^2}\Bigl( 2\frac{E}{w} - 3\frac{b^2}{w^2}\Bigr)\right\rbrack \; ,
\lab{22.a} \\
&
r_s = 2Gw\left\lbrack 1 + 
\frac{4b^2}{m_w^2}\Bigl( 2\frac{E}{w} - 3\frac{b^2}{w^2}\Bigr)\right\rbrack \; ,
\;\; C = - \frac{8\pi G}{w} ; .
\nonu
% \lab{22.b}
\end{align}

Thus, we end up with the following $x$-space expression for the metric tensor:
\be
g_{00}= - \Bigl( 1 - \frac{r_t}{r}\Bigr) \; ,\;\; g_{0i}=0 \; ,\;\;
g_{ij} = \delta_{ij} + r_s \frac{x_i x_j}{r^3} 
- \frac{8\pi G}{w} \delta_{ij} \delta(\mathbf{x}) \; .
\lab{23}
\ee
The last ($\delta$-function) term does not contribute to the classical motion
and will be ignored in the sequel (it may only be relevant, for a 
quantum s-wave effect). Clearly, in the text body limit, i.e., 
for $(m_1 + m_2)^2 \gg m_1 m_2$, and for $|w(m_1 + m_2)^{-1} - 1| \ll 1$
(slow motion), the right-hand sides of Equations (23) go into the 
linearized Schwarzschild solution (both $r_t$ and $r_s$ tending to the 
Schwarzschild radius $2(m_1 + m_2)G$).

\bigskip

{\bf 6.} We are now prepared to treat the classical gravitational 
two-body problem by inserting the metric (23) into the Hamiltonian constraint
(7). Going to spherical co-ordinates, we can rewrite Equation (7) in the form
\be
H= \frac{1}{2}\left\lbrack m_w^2 - \Bigl( 1 - \frac{r_t}{r}\Bigr)^{-1} p_0^2
+ \Bigl( 1 - \frac{r_s}{r}\Bigr)p_r^2 + \frac{1}{r^2} p_{\theta}^2
+ \frac{1}{\sin^2 \theta} p_{\varphi}^2 \right\rbrack \approx 0
\lab{24}
\ee
A standard computation using the initial condition 
$\theta = \pi/2, p_\theta = 0$ (cf. [17] ) gives:
\be
-p_0 = E \;({\stackrel{.}{E}}=0)\;\;, \;\; p_{\varphi} = J \;
({\stackrel{.}{J}}=0)\;\;,\;\;
p_r^2 \Bigl( 1 - \frac{r_t}{r}\Bigr) = 
b^2 + E^2 \frac{r_t}{r}\Bigl( 1 - \frac{r_t}{r}\Bigr)^{-1} - \frac{J^2}{r^2}
\; .
\lab{25}
\ee
Introducing the radius variable $u = r^{-1}$ and setting 
$du/d\varphi = u^{\prime}$ we obtain
\be
J^2 (u^{\prime\, 2} + u^2 - r_s u^3) + 
\lbrack r_s b^2 - r_t E^2(1-r_s u)(1-r_t u)^{-1}\rbrack u^{\prime} = b^2 \; . 
\lab{26}
\ee
We look for a solution of this equation of the form
\be
u= \ell^{-1}\bigl\lbrack 1 + 
\epsilon (\cos \eta\varphi + f(\eta\varphi))\bigr\rbrack \; ,
\lab{27}
\ee
where the natural dimensionless small parameter is now $r/\ell$ . 
The unknown function $f(\varphi)$ is expected to be a small correction 
(of order $r/\ell$) to the Schwarzschild-like solution. Inserting in 
(26) and comparing the coefficients of $\cos 2\eta\varphi,\,\cos \eta\varphi$,
and the constant term, we find:
\begin{align}
&
\eta = 1 - \frac{3r_t}{2\ell} + \frac{r_t - r_s}{2\ell}\;\;, \quad
\epsilon^2 = 1 + \frac{4r_t}{\ell} +
\frac{\ell^2 b^2}{J^2}\Bigl( 1 + \frac{3r_t + r_s}{\ell}\Bigr) \; ,
\lab{28} \\
&
\ell = 2J^2 (r_t m_w^2 - r_s b^2)^{-1} + O(r_s) \; .
\nonu 
\end{align}
The terms containing $f, f^{\prime}$ and $\cos^3 \eta\varphi$ lead to 
the differential equation 
\be
\sin \eta\varphi \cdot f^{\prime} - \cos \eta\varphi \cdot f
+\frac{\epsilon r_s}{2\ell} \cos^3 \eta\varphi = 0 \; .
\lab{29}
\ee
Its solution, satisfying $f < |\cos \eta\varphi|$ for all $\varphi$ is
\be
f = \frac{\epsilon r_s}{2\ell}\bigl( 1 - |\sin \eta\varphi|\bigr)^2 
+ O\Bigl\lbrack \left(\frac{r_s}{\ell}\right)^2 \Bigr\rbrack 
\lab{30}
\ee
(which is of order $r_s/\ell$ in accord with our expectation).

The solution (27), (28) and (30) so obtained, reduces to the classical 
one [1] in the semirelativistic and test body limit.\footnote{This is to 
be contrasted with the results of some previous first-order (in G) 
relativistic approaches [18], which give incorrect values for $\eta$ even 
in the test body limit.}
It is expected to give a more accurate description of the two-particle 
motion for relativistic velocities and weak gravitational forces.

\bigskip

ACKNOWLEDGEMENTS

The authors would like to thank Professor Abdus Salam and Professor 
P. Budinich, the International Atomic Energy Agency and UNESCO for 
hospitality at the International Centre for Theoretical Physics, Trieste. 
Two of them (E.R.N. and I.T.T.) acknowledge financial support from 
the Scuola Internazionale Superiore di Studi Avanzati, Trieste. 
We also thank the referee for his constructive criticism which led to 
an improvement in the final version of the paper.

\bigskip

 REFERENCES

{\bf 1.} Einstein, A., Infeld, L., and Hoffmann, B., Ann. Math. 39, 65 (1938); 
 41, 455 (1940); \\ 
   Infeld, L. and Plebanski, J., Motion and Relativity, Pergamon Press, 
   Oxford and PWN,  Warszawa, 1960 (see, in particular, Section 3 of Chapter V).
   
{\bf 2.} Infeld, L. and Michalska-Trautman, R., Ann. Phys. (N. Y ) 55, 561 (1969).
 
{\bf 3.} Petrova, N.M., Zh. Eksp. Teor. Fiz. 19, 989 (1949) (in Russian).
 
{\bf 4.} Todorov, I.T.,Phys. Rev. D3, 2351 (1971); see also Properties of 
 Fundamental Interactions, ed. A. Zichichi, Editrice Compositori, Bologna, 
 1973, Vol. 9C, pp. 951-979.
   
{\bf 5.} Rizov, V.A. and Todorov, I.T., Soy. J. Part. Nucl. 6, 269 (1975);
   Rizov, V.A., Todorov, I.T., and Aneva, B.L.,Nuc/. Phys. B98, 447 (1975).
   
{\bf 6.} Logunov, A.A. and Tavkhelidze, A.N.,Nuovo Cimento 29, 380 (1963);
   Logunov, A.A. et al., Nuovo Cimento 30, 134 (1963).
   
{\bf 7.} Todorov, I.T., 'Dynamics of Relativistic Point Particles as a Problem
 with Constraints', Commun. JINR E2-10125, Dubna (1976); \\
   Molotkov, V.V. and Todorov, I.T., 'Gauge Dependence of World Lines
   and Invariance of the    S-matrix in Relativistic Point Particle Dynamics',
   ICTP, Trieste, Internal Report 1C/80/101; 
   Commun. Math. Phys. 79, 111 (1981).
   
{\bf 8.} Todorov, I.T., 'Constraint Hamiltonian Dynamics of Directly Interacting 
 Relativistic Point Particles', Lectures presented at the Summer School in 
 Mathematical Physics, Bogazici University, Bebek, Istanbul (1979) and at 
 the XVIIth Winter School of Theoretical Physics, Karpacz (1980) 
 (to be published).
   
{\bf 9.} Droz-Vincent, Ph., Ann. Inst. H Poincare 27, 407 (1977); 
 'N-body Relativistic Systems',    College de France preprint, Paris (1979);\\
   Bel, L., and Martin, J., Ann. Inst. IL PoincaW 22A, 173 (1975); \\
   Fustero, F.X. and Lapiedra, R., Phys. Rev. D17, 2821 (1978).
   
{\bf 10.} Dominici, D., Gomis, J., and Longhi, G., Nuovo Cimento 48A, 257 (1978); 
48B, 152 (1978); \\
   Giachetti, R. and Sorace, E. 'Relativistic Two-Body Interactions: 
   A Hamiltonian Formulation'    (to appear in Nuovo Cimento).
   
{\bf 11.} Takabayasi, T., 'Relativistic Mechanics of Confined Particles as 
Extended Model of Hadrons -- The Bifocal Case', Prog. Theor. Phys. Suppl. 
No. 67,1 (1979).

{\bf 12.} Gupta, S. and Radford, S., Phys. Rev. D21, 2213 (1980) and 
references therein.

{\bf 13.} Gupta, S., Proc. Phys. Soc. 65A, 161, 608 (1952); \\
   Corinaldesi, E., Proc. Phys. Soc. 69A, 189 (1956).
   
{\bf 14.} Dirac, P.A.M., Lectures on Quantum Mechanics, Belfer Grad. School of Science, Yeshiva 
   Univ., N.Y., 1964.
   
{\bf 15.} Penrose, R., in Relativity, Groups and Topology, eds. C.M. de Witt, and
B. de Witt, Gordon and Breach, N.Y., 1964.

{\bf 16.} Weinberg, S., in Gravitational Theories Since Einstein, eds. 
S. Hawking and W. Israel,   Cambridge Univ. Press, Cambridge, 1979.

{\bf 17.} Weinberg, S., Gravitation and Cosmology, Wiley, N.Y., 1972.

{\bf 18.} Havas, P. and Goldberg, J., Phys. Rev., 128, 398 (1962).

\bigskip

(Received December 12, 1980; revised version April 23, 1981)

\footnotesize
 \bibliography{/Users/vani/Files/PAPERS/PAPERSLIBRARY/ABib}
  \bibliographystyle{/Users/vani/Files/PAPERS/PAPERSLIBRARY/utcaps}

% \bibliography{ABib}
% \bibliographystyle{utcaps}

% \end{document}

\end{document}